\documentclass{article}
\usepackage[utf8]{inputenc}

\usepackage{amsfonts}
\usepackage{amsmath}
\usepackage{amssymb}
\usepackage{amsthm}
\usepackage{mathtools}
\newcommand{\indep}{\perp \!\!\! \perp}

\usepackage{amsmath}
\usepackage{graphicx}
\usepackage{enumerate}
\usepackage{natbib}
\usepackage{url} 
\usepackage{amssymb, amsthm, bbm}
\usepackage{caption,subcaption,placeins,color,dsfont}
\usepackage{fancyhdr}
\usepackage{hyperref}
\usepackage{accents}
\usepackage{mathtools}
\usepackage{algorithm}
\usepackage{algcompatible}
\usepackage{multirow}
\usepackage[table,x11names]{xcolor}
\usepackage{booktabs}

\DeclareMathOperator*{\argmin}{arg\,min}

\title{Estimating Treatment Effects Using Observational Data and Experimental Data with Non-overlapping Support}
\author{Kevin Han}
\date{January 2021}

\begin{document}

\maketitle

\section{Introduction}

When estimating treatment effects, the golden standard is to conduct a randomized experiment and then contrast outcomes associated with the treatment group and the control group. However, in many cases, randomized experiments are either conducted with a much smaller scale compared to the size of the target population or accompanied with certain ethical issues and thus hard to implement. Therefore, researchers usually rely on observational data to study causal connections. The downside is that the unconfoundedness assumption, the key to validate the use of observational data is hard to verify and almost always violated. Hence, any conclusion drawn from observational data should be further analyzed with great care. Given the richness of observational data and usefulness of experimental data, researchers hope to develop credible method to combine the strength of the two. In this paper, we consider a setting where the observational data contain the outcome of interest as well as a surrogate outcome while the experimental data contain only the surrogate outcome. We propose a simple estimator to estimate the average treatment effect of interest using both the observational data and the experimental data.

The remainder of the paper is organized as follows. Section~\ref{sec: setup} introduces the basic setup. In Section~\ref{sec: estimator} we develop our method to estimate the treatment effect of primary outcome by using information from the experimental study. In Section~\ref{sec: applications}, we discuss several widely-studied extensions to the basic setup and give concrete solution to each extension. Section~\ref{sec: simulations} compare several different methods through simulations.

\section{Setup}
\label{sec: setup}

Suppose we want to estimate the treatment effect of an intervention on some primary outcome $Y^P \in \mathbb{R}$. For each unit $i$ in the observational study, along with the treatment assignments $W_i$, the outcome $Y_i^P$, we also observe another surrogate outcome $Y_i^S \in \mathbb{R}$ and record a number of pre-treatment covariates $X_i$. Here, the surrogate outcome $Y^S$ can be any variable that change after treatment. In this paper, we mainly discuss the case that $Y^S$ is one-dimensional, but our method can be generalized naturally to multi-dimensional surrogate outcome. If the unconfoundedness assumption is satisfied, i.e.
\begin{equation*}
    Y_i(1), Y_i(0) \indep W_i | X_i,
\end{equation*}
then either IPW estimator or AIPW estimator suffices for our propose. However, there exist many settings under which researchers do not believe unconfoundedness holds, hence makes estimating treatment effect using only the observational study impossible. To this end, we assume that there is another prior study on the surrogate outcome $Y^S$ such that unconfoundedness holds. Typically, this can be a small-scale randomized experiment on the surrogate outcome. In summary, we assume that we have two samples: the observational sample and the experimental sample. There is a $(X_i, W_i, Y_i^S, Y_i^P)$ tuple associated with every unit $i$ in our observational sample and a $(X_i, W_i, Y_i^S)$ tuple associated with every unit $i$ in the experimental sample. The experimental sample size $N_E$ is considered to be much smaller than the observational sample size $N_O$. We are interested in the quantity
\begin{equation*}
    \tau^P = \mathbb{E}[Y_i^P(1) - Y_i^P(0) | G_i = O],
\end{equation*}
where $G_i$ is the indicator function of which sample unit $i$ belongs to. We note by passing that this is exactly the same setup as in \cite{athey2020}.

\section{A simple estimator}
\label{sec: estimator}

We develop our simple estimator in this section. To be able to point-identify the ATE of $Y^P$, we assume the following structural model of $Y^P$:
\begin{equation}
    Y_i^P = f(X_i, Y_i^S, \epsilon_i), \qquad \epsilon_i \indep X_i, Y_i^S, \label{yp}
\end{equation}
i.e. all the effect of treatment on the primary outcome is mediated through the surrogate outcome. Therefore, the surrogate outcome together with pre-treatment covariates determine the primary outcome. Now, $\tau^P$ is identifiable. 

To see this, define
\begin{equation*}
    \tau^S(x) = \mathbb{E}[Y_i^S(1) - Y_i^S(0) | X_i = x]
\end{equation*}
and
\begin{equation*}
    \mu(x, y) = \mathbb{E}\left[Y_i^P|X_i = x, Y_i^S = y, G_i = O\right].
\end{equation*}
then $\mathbb{E}\left[Y_i^P(w)\right]$ is just $\mathbb{E}[\mu(X_i, Y_i^S(w))]$. The joint distribution of $X_i$ and $Y_i^S(w)$ is identifiable from the experimental sample because of unconfoundedness. There is a concrete model that we know well: $Y_i^P = \rho Y_i^S + f(X_i) + \epsilon_i$ where $\epsilon_i$ is independent with $Y_i^S$ and $X_i$. For such model, we can use Robinson residual-in-residual method to estimate $\rho$ and the final estimate of the ATE would be consistent. For the general case, we can estimate the $\tau^P$ as follows:
\begin{enumerate}
    \item Regress $Y^P$ on $Y^S$ and $X$ to obtain an estimate of $\mu$, $\hat{\mu}$.
    \item Estimate the conditional average treatment effect $\tau(x)$ on the surrogate outcome $Y^S$, obtain an estimate of $\tau$, $\hat{\tau}$.
    \item Define $\hat{Y}_i^S(1) = Y_i^S$ if $W_i = 1$ and $\hat{Y}_i^S(1) = Y_i^S + \hat{\tau}(X_i)$ if $W_i = 0$.
    \item Now we can estimate $\mathbb{E}\left[Y_i^P(1)\right]$ by $\frac{1}{N_O}\sum_{i = 1}^{N_O}\hat{\mu}(X_i, \hat{Y}_i^S(1))$.
    \item Define $\hat{Y}_i^S(0) = Y_i^S$ if $W_i = 0$ and $\hat{Y}_i^S(0) = Y_i^S - \hat{\tau}(X_i)$ if $W_i = 1$.
    \item Now we can estimate $\mathbb{E}\left[Y_i^P(0)\right]$ by $\frac{1}{N_O}\sum_{i = 1}^{N_O}\hat{\mu}(X_i, \hat{Y}_i^S(0))$.
    \item The final estimate would be $\hat{\tau}^P = \frac{1}{N_O}\sum_{i = 1}^{N_O}\hat{\mu}(X_i, \hat{Y}_i^S(1)) - \frac{1}{N_O}\sum_{i = 1}^{N_O}\hat{\mu}(X_i, \hat{Y}_i^S(0))$.
\end{enumerate}
With the above procedure, to estimate the ATE on the primary outcome, we only need one model for the conditional response function $\mu$ and one model for CATE estimation. In the next section, we will discuss different variants of the procedure above in different scenarios.

\section{Applications}
\label{sec: applications}
In the previous section, we develop a general procedure to combine both the experimental sample and the observational sample. It relies on first estimating the conditional average treatment effect on the surrogate outcome and then correcting the surrogate outcomes in the observational sample. Estimating the conditional agverage treatment effect (CATE) is usually a case-by-case problem and involves different estimation methods for different settings. In this section, we discuss four settings where we can apply the estimator in Section~\ref{sec: estimator} with different versions of step 2. We will also discuss the setting where we drop the unconfoundedness assumption on experimental sample. In fact, as long as the conditional average treatment effect $\tau$ is identifiable, unconfoundedness is not necessary.

\subsection{Different support of pre-treatment covariates}
The first scenario that we consider is the setting in \cite{kallus2018} where the support of pre-treatment covariates in the experimental sample is different from the support of pre-treatment covariates in the observational sample. This is usually the case in practice since the experimental sample typically comes from historical data and we cannot guarantee that the experimental study and the observational study are targeting exactly the same population. Under this setting, if we only use the experimental sample to estimate the conditional average treatment effect, we need to extrapolate on the observational sample. Such extrapolation will be more problematic if the sample size of the experimental sample is much smaller compared to the sample size of the observational sample. Therefore, for our propose, we should calibrate our conditional average treatment estimate on the experimental sample. \cite{kallus2018} noticed that if we define $e^E(x) = \mathbb{P}(W_i = 1 | X_i = x, G_i = E)$ and $q^E(X_i) = \frac{W_i}{e^E(X_i)} - \frac{1 - W_i}{1 - e^E(X_i)}$, then
\begin{equation*}
    \mathbb{E}[q^E(X_i)Y_i | X_i] = \tau(X_i).
\end{equation*}
Define $\omega(x)$ to be $\mathbb{E}[Y_i | W_i = 1, X_i = x, G_i = O] - \mathbb{E}[Y_i | W_i = 0, X_i = x, G_i = O]$, then the above observation motivates the following procedure to estimate the conditional average treatment effect of the surrogate outcome on the observational sample:
\begin{enumerate}
    \item Run any CATE algorithm on the observational sample, obtain $\hat{\omega}$.
    \item Solve the following optimization problem to obtain $\hat{\theta}$:
    \begin{equation*}
        \hat{\theta} = \argmin_\theta\sum_{i = 1}^{N_E}(q^E(X_i)Y_i - \hat{\omega}(X_i) - \theta^TX_i)^2
    \end{equation*}
    \item $\hat{\tau}(x) = \hat{\theta}^T x + \hat{\omega}(x)$.
\end{enumerate}
Now we can use the above estimate of $\tau$ for the estimator described in Section~\ref{sec: estimator}. Essentially, the idea here is to use a loss function to estimate the difference between ill-posed target $\omega$ and the true quantity of interest $\tau$. A more general version can be obtained if we do not fit a linear function but a non-parametric function of $X_i$.

\subsection{IV setting in the experimental sample}
In this section, we drop our unconfoundedness assumption on the experimental sample and consider the instrumental variable setting which is widely-studied in econometrics literature.
\subsubsection{Constant effect}
We start with the simplest instrumental variable setting where the effect is constant. In particular, we consider a setting where in the experimental sample we have an instrumental variable $Z$ with the following structural model:
\begin{equation*}
    Y_i^S = \alpha^TX_i + W_i\tau + \epsilon_i, \qquad \epsilon_i \indep Z_i \label{ys}
\end{equation*}
\begin{equation*}
    W_i = \beta^TX_i + Z_i\gamma + \xi_i. \label{w}
\end{equation*}
Such model is introduced in almost every econometrics textbook, for example, in \cite{angrist09}. It can be seen easily that the parameter $\tau$ is exactly the conditional average treatment effect of $Y^S$. It is well known that we can then estimate it by two-stage least squares (2SLS) in usual instrumental variable literature. 

\subsubsection{Nonparametric IV}
Now we consider a more general instrumental variable setting. Specifically, we consider the following model:
\begin{equation*}
    Y_i^S = \tau(X_i)W_i + g(X_i) + \epsilon_i, \qquad \epsilon_i \indep Z_i
\end{equation*}
This is a special case of the more general nonparametric instrumental variable model \citep{newey2003, hall2005, horowitz2011}. Here, to estimate $\tau$, we can follow \citep{hall2005}. First, note that
\begin{align*}
    \mathbb{E}[Y|W = 1, Z = z] &= \mathbb{E}[\tau(X)| W = 1, Z = z] + \mathbb{E}[g(X) | W = 1, Z = z] \\
    &= \int_0^1(\tau(x) + g(x))f_{X|W = 1, Z}(x, z)dx \\
    &= \int_0^1(\tau(x) + g(x))\frac{f_{XZ|W = 1(x, z)}}{f_{Z|W = 1}(z)}dx \\
\end{align*}
Therefore,
\begin{equation*}
    \mathbb{E}[Y|W = 1, Z = z]f_{Z|W = 1}(z) = \int_0^1(\tau(x) + g(x))f_{XZ|W = 1}(x, z)dx,
\end{equation*}
so
\begin{equation}
    \mathbb{E}[Y|W = 1, Z = z]f_{Z|W = 1}(z)f_{XZ|W = 1}(u, z) = \int_0^1(\tau(x) + g(x))f_{XZ|W = 1}(x, z)f_{XZ|W = 1}(u, z)dx \label{equality}
\end{equation}
If we define
\begin{equation*}
    t(x, u) = \int_{0}^{1}f_{XZ|W = 1}(x, z)f_{XZ|W = 1}(u, z)dz
\end{equation*}
and integrate both sides of (\ref{equality}) with respect to $z$, then we have
\begin{equation*}
    \mathbb{E}[Yf_{XZ|W = 1}(u, Z)] = \int_0^1(\tau(x) + g(x))t(x, u)dx
\end{equation*}
for any $u \in [0, 1]$ where the expectation on left hand side is taken with respect to the conditional joint distribution $(Y, Z | W = 1)$. If we define
\begin{equation*}
    (Th)(u) = \int_0^1h(x)t(x, u)dx
\end{equation*}
and
\begin{equation*}
    r(u) = \mathbb{E}[Yf_{XZ|W = 1}(u, Z)]
\end{equation*}
then we arrive at the following operator equation
\begin{equation*}
    r(u) = (T(\tau + g))(u).
\end{equation*}
We can estimate $\tau + g$ using Hall-Horowitz estimator. Similarly, we have another operator equation where we only have $g$ by conditioning on $W = 0$. With that equation, we are able to estimate $g$. Then we can estimate $\tau$ by taking the difference. 

\cite{hall2005} give good theoretical properties of this method. However, it involves estimating density functions which is unstable in practice. In fact, \cite{hall2005} aims to address the general nonparametric IV problem while we only care about $\tau(x)$. 

With our structural model assumption, \cite{athey2019} propose the Generalized Random Forests (GRF) to estimate the conditional average treatment effect $\tau$. We recommend to use GRF for estimating $\tau$. In fact, one advantage of using GRF is that it can be generalized to the setting where $W$ is no longer binary but a real number.

\subsection{IV setting with different support of pre-treatment covariates}
In this section, we combine our two considerations above. We want to address the setting where we have different support of pre-treatment covariates and a nonparametric instrumental variable model for the experimental sample. We first note that if we let
\begin{equation*}
    \mu(x) = \mathbb{E}[Y|X = x]
\end{equation*}
\begin{equation*}
    \pi(x) = \mathbb{E}[Z|X = x]
\end{equation*}
\begin{equation*}
    e(x) = \mathbb{E}[W|X = x]
\end{equation*}
\begin{equation*}
    m(x) = \mathbb{E}[YZ|X = x]
\end{equation*}
\begin{equation*}
    \gamma(x) = \mathbb{E}[WZ|X = x]
\end{equation*}
Then
\begin{equation*}
    \tau(x)[\gamma(x) - e(x)\pi(x)] - [m(x) - \mu(x)\pi(x)] = 0.
\end{equation*}
Therefore, we can write
\begin{equation*}
    \tau(x) = \argmin_{\tau: \mathcal{X} \rightarrow \mathbb{R}}\mathbb{E}[(\tau(x)[\gamma(x) - e(x)\pi(x)] - [m(x) - \mu(x)\pi(x)])^2].
\end{equation*}
It is possible to directly estimate $\tau$ with the above loss function but we found that it does not work well when we have multi-dimensional pre-treatment covariates as we need to estimate many nuisance parts and the errors may aggregate. However, this loss defining property of $\tau$ motivates the following procedure (which we abbreviate by Kallus IV):
\begin{enumerate}
    \item Run any CATE estimation algorithm $\mathcal{Q}$ on $\{X_i, W_i, Y_i^S\}_{i = 1}^{m}$ to get an estimate $\hat{\omega}$.
    \item Solve the following optimization algorithm on the experimental sample:
    \begin{equation*}
        \hat{\theta} = \argmin_{\theta}\sum_{i = 1}^{n}\left([\hat{m}(x_i)- \hat{\mu}(x_i)\hat{\pi}(x_i)] - (\theta^Tx_i + \hat{\omega}(x_i))\times[\hat{\gamma}(x_i) - \hat{e}(x_i)\hat{\pi}(x_i)]]\right)^2
    \end{equation*}
    \item Use $\hat{\omega}(x) + \hat{\theta}^Tx$ as our estimate of CATE on the surrogate.
\end{enumerate}
Essentially we are adapting the procedure in \cite{kallus2018} with a different objective function when estimating $\theta$. Similar to our remark in the unconfounded case, we can actually fit a non-parametric function of $X_i$ instead of a linear function. However, we found that this will give us rather unstable estimates when we have many covariates.

\section{Simulations}
\label{sec: simulations}
In the previous sections, we outlined a procedure to estimate the average treatment effect of the primary outcome given prior information in the experimental sample and considered three scenarios in which we can utilize our procedure described in Section~\ref{sec: estimator}. In this section, we compare several estimators through simulations. In particular, we hope to compare our procedure with the canonical imputation estimator in \cite{athey2020} when we have an unconfounded experimental sample.

We consider two settings: there is no confounding in the experimental sample (i.e., we have either a randomized experiment or an unconfounded experiment) and there is confounding (we assume a nonparametric IV model for the experimental sample). For each setting, we consider two subcases: the support of the pre-treatment covariates in the experimental sample is the same as the support of pre-treatment covariates in the observational sample and the support of the pre-treatment covariates in the experimental sample is not the same as the support of pre-treatment covariates in the observational sample (but they do overlap). When there is no confounding, we compare three estimators: the imputation estimator in \cite{athey2020}, our estimator with $\tau(x)$ estimated by generalized random forest and our estimator with $\tau(x)$ estimated by the approach in \cite{kallus2018}. When there is confounding, both the imputation estimator and the approach in \cite{kallus2018} are no longer valid as they require the experimental sample to be unconfounded. Hence, we will compare two estimators: our estimator with $\tau(x)$ estimated by generalized random forest and our estimator with $\tau(x)$ estimated by Kallus IV.

We work with the following data generating mechanism:
\begin{equation*}
    X_i \sim \mathcal{N}(0, I_{p \times p}), \qquad \epsilon_i \sim \mathcal{N}(0, 1), \qquad Z_i \sim Binom(1/3),
\end{equation*}
\begin{equation*}
    Q_i \sim Binom(1/(1 + e^{-\omega\epsilon_i})), \qquad W_i = Z_i \wedge Q_i,
\end{equation*}
\begin{equation*}
    Y_i^S = \mu(X_i) + (W_i - 1/2)\tau(X_i) + \epsilon_i.
\end{equation*}
and
\begin{equation*}
    Y_i^P = \sum_{j = 1}^{\kappa}X_i^{(j)} + (X_i^{(p)})^2 + 2Y_i^S + (X_i^{(p-2)} + X_i^{(p-1)}X_i^{(p-3)})Y_i^S + \xi_i
\end{equation*}
i.e., $Y^P = f(Y^S, X, \xi)$ where $\xi$ is independent noise. This is the same setting as in the appendix of \cite{athey2019}.

Now, we can adjust several parameters in the data generating mechanism to satisfy different conditions. 
\begin{enumerate}
    \item \textbf{Presence of confounding}: we vary $\omega$ to be either 0 or 1. If $\omega = 0$, there is no confounding, otherwise there is confounding and we are in the nonparametric IV model.
    \item \textbf{Sparsity of the signal}: $\kappa_\tau \in \{2, 4\}$.
    \item \textbf{Additivity of the signal}: When true, $\tau(x) = \sum_{j = 1}^{\kappa_\tau}\max\{0, x_j\}$; when false, $\tau(x) = \max\{0, \sum_{j = 1}^{\kappa_\tau}x_j\}$.
    \item \textbf{Presence of nuisance terms}: When true, $\mu(x) = 3\max\{0, x_5\} + 3\max\{0, x_6\}$ or $\mu(x) = 3\max\{0, x_5 + x_6\}$ depending on the additive signal condition; when false, $\mu(x) = 0$.
    \item \textbf{Identical support}: When true, we assume the distribution of the covariates in the experimental sample and that in the observational sample are the same; when false, $X_i \sim \mathcal{N}([1, \cdots, 1]^T, I_{p\times p})$ in the observational sample.
\end{enumerate}
Here, we fix the dimension of $X_i$, $p$ to be 10, the experimental sample size, $n$ to be 300 and the observational sample size, $m$ to be 1000. We are interested in the treatment effect on $Y^P$. We compare different methods based on mean squared erro (MSE). To calculate MSE, we use Monte Carlo method to estimate the true value of ATE and generate 200 realizations.

\begin{table} 
\centering
\scalebox{0.8} {
\begin{tabular}{c l c c c c c c c c c} 
\toprule 
$\omega$ & $\kappa_\tau$ & Additivity & Nuisance & Identical Support & MC Estimate & GRF & Imputation & Kallus & Winner\\
\midrule
    \multirow{1}{*}{0} 
             & 2 & Yes & Yes & Yes & 1.62 & 0.19 & 1.02 & 0.34 & GRF\\
            \midrule
    \multirow{1}{*}{0} 
             & 2 & Yes & No & Yes & 1.58 & 0.12 & 0.22 & 0.24 & GRF\\
            \midrule
    \multirow{1}{*}{0} 
             & 4 & No & Yes & Yes & 2.10 & 0.22 & 1.13 & 0.55 & GRF\\
            \midrule
    \multirow{1}{*}{0} 
             & 4 & No & No & Yes & 2.10 & 0.14 & 0.26 & 0.41 & GRF\\
            \midrule
    \multirow{1}{*}{0} 
             & 2 & Yes & Yes & No & 8.73 & 30.91 & 43.38 & 72.83 & GRF\\
            \midrule
    \multirow{1}{*}{0} 
             & 2 & Yes & No & No & 8.67 & 27.28 & 36.00 & 6.45 & Kallus\\
            \midrule
    \multirow{1}{*}{0} 
             & 4 & No & Yes & No & 8.11 & 18.56 & 29.07 & 51.25 & GRF\\
            \midrule
    \multirow{1}{*}{0} 
             & 4 & No & No & No & 8.11 & 15.98 & 23.29 & 7.05 & Kallus\\
\bottomrule 
\end{tabular}}
\caption{Simulation results for $\omega = 0$} 
\label{table:omega=0} 
\end{table}

\begin{table} 
\centering
\scalebox{0.9} {
\begin{tabular}{c l c c c c c c c c c} 
\toprule 
$\omega$ & $\kappa_\tau$ & Additivity & Nuisance & Identical Support & MC Estimate & GRF & Kallus IV & Winner\\
\midrule
    \multirow{1}{*}{1} 
             & 2 & Yes & Yes & Yes & 1.63 & 0.46 & 0.65 & GRF\\
            \midrule
    \multirow{1}{*}{1} 
             & 2 & Yes & No & Yes & 1.55 & 0.26 & 0.80 & GRF\\
            \midrule
    \multirow{1}{*}{1} 
             & 4 & No & Yes & Yes & 2.12 & 0.49 & 0.64 & GRF\\
            \midrule
    \multirow{1}{*}{1} 
             & 4 & No & No & Yes & 2.11 & 0.30 & 0.51 & GRF\\
            \midrule
    \multirow{1}{*}{1} 
             & 2 & Yes & Yes & No & 8.73 & 31.60 & 28.27 & Kallus IV\\
            \midrule
    \multirow{1}{*}{1} 
             & 2 & Yes & No & No & 8.72 & 27.84 & 10.80 & Kallus IV\\
            \midrule
    \multirow{1}{*}{1} 
             & 2 & No & Yes & No & 6.35 & 15.00 & 31.79 & GRF\\
            \midrule
    \multirow{1}{*}{1} 
             & 2 & No & No & No & 6.33 & 12.64 & 11.01 & Kallus IV\\
            \midrule
    \multirow{1}{*}{1} 
             & 4 & No & Yes & No & 8.11 & 17.93 & 32.65 & GRF\\
            \midrule
    \multirow{1}{*}{1} 
             & 4 & No & No & No & 8.09 & 15.35 & 16.72 & GRF\\
            \midrule
    \multirow{1}{*}{1} 
             & 4 & Yes & No & No & 17.30 & 109.78 & 28.89 & Kallus IV\\
            \midrule
    \multirow{1}{*}{1} 
             & 4 & Yes & Yes & No & 17.38 & 114.74 & 42.00 & Kallus IV\\
\bottomrule 
\end{tabular}}
\caption{Simulation results for $\omega = 1$} 
\label{table:omega=1} 
\end{table}

Table \ref{table:omega=0} and \ref{table:omega=1} show the simulation results. We see that when we have identical support of pre-treatment covariates, GRF performs better than the other two methods regardless of confounding issue. This makes sense since when the support does not change, we do not actually need to extrapolate, hence the Kallus method won't improve much. When the support is different, generally Kallus and Kallus IV are also competitive. In fact, when there is confounding, Kallus IV performs better than GRF.

To further investigate the case of different support, we change the above setting slightly. Now we assume that when the support is not identical, the support of pre-treatment covariates of the experimental sample will be contained in the support of pre-treatment covariates of the observational sample (instead of just overlap). Specifically,
\begin{enumerate}
    \item[5a] \textbf{Identical support}: When true, we assume the distribution of the covariates in the experimental sample and that in the observational sample are the same: $X_i^{(j)} \sim $ Uniform$(-1, 1)$; when false, $X_i^{(j)} \sim $ Uniform$(-1, 1)$ in the experimental sample and $X_i \sim \mathcal{N}(0, I_{p\times p})$ in the observational sample.
\end{enumerate}

\begin{table} 
\centering
\scalebox{0.8} {
\begin{tabular}{c l c c c c c c c c c} 
\toprule 
$\omega$ & $\kappa_\tau$ & Additivity & Nuisance & Identical Support & MC Estimate & GRF & Kallus/Kallus IV & Winner\\
\midrule
    \multirow{1}{*}{0} 
             & 2 & Yes & Yes & No & 1.61 & 0.60 & 0.30 & Kallus\\
            \midrule
    \multirow{1}{*}{0} 
             & 2 & Yes & No & No & 1.60 & 0.58 & 0.29 & Kallus\\
            \midrule
    \multirow{1}{*}{0} 
             & 4 & No & Yes & No & 2.11 & 1.12 & 0.54 & Kallus\\
            \midrule
    \multirow{1}{*}{0} 
             & 4 & No & No & No & 2.10 & 1.16 & 0.45 & Kallus\\
            \midrule
    \multirow{1}{*}{1} 
             & 2 & Yes & Yes & No & 1.60 & 0.77 & 0.71 & Kallus IV\\
            \midrule
    \multirow{1}{*}{1} 
             & 2 & Yes & No & No & 1.60 & 0.70 & 0.68 & Kallus IV\\
            \midrule
    \multirow{1}{*}{1} 
             & 2 & No & Yes & No & 1.34 & 0.61 & 0.83 & GRF\\
            \midrule
    \multirow{1}{*}{1} 
             & 2 & No & No & No & 1.35 & 0.58 & 0.71 & GRF\\
            \midrule
    \multirow{1}{*}{1} 
             & 4 & No & Yes & No & 2.10 & 1.21 & 0.66 & Kallus IV\\
            \midrule
    \multirow{1}{*}{1} 
             & 4 & No & No & No & 2.08 & 1.24 & 0.57 & Kallus IV\\
            \midrule
    \multirow{1}{*}{1} 
             & 4 & Yes & No & No & 3.21 & 2.37 & 0.60 & Kallus IV\\
            \midrule
    \multirow{1}{*}{1} 
             & 4 & Yes & Yes & No & 3.23 & 2.26 & 0.54 & Kallus IV\\
\bottomrule 
\end{tabular}}
\caption{Simulation results, inclusion of the support} 
\label{table:omega=1, beta} 
\end{table}
Table \ref{table:omega=1, beta} shows the simulation results. We see that similar to the simulation results in the previous two tables, Kallus/Kallus IV performs better than GRF when we have different support.

\section{A real data example}
In this section, we investigate the performance of our procedure on a real dataset. We utilize the famous Tennessee STAR study \citep{DVN/SIWH9F_2008}. This dataset is also used in \cite{kallus2018} and \cite{athey2020}. We use it in a different manner. Specifically, we select the following covariates for each student: gender, race, birth month, birthday, birth year, free lunch given or not, teacher id, student home location. We focus on two outcomes: average grade in year 1 and average grade in year 3. We remove all the records with missing outcome variables. Now, in this study, the treatment is whether or not the student is in small class (treatment) or regular class (control). After cleaning the data, we have a dataset with 2498 units, 9 covariates, 1 treatment variable and 2 outcome variables. We use the method in \cite{athey2020using} to generate a large population, which we view as the ground truth. We call this ground truth dataset $\mathcal{D}_{gt}$. To assess different methods, we do the following:
\begin{enumerate}
    \item Use $\mathcal{D}_{gt}$ to calculate the average treatment effect of average grade in year 3. This estimate $\tau_{gt}$ will be viewed as the ground truth.
    \item Repeat the following steps 500 times.
    \item Sample $n_{\text{exp}}$ rural or inner-city students together with all the covariates except the student location covariate, treatment variable and average grade in year 1. This is our experimental sample $\mathcal{D}_E$.
    \item Sample $n_{\text{obs}}/4$ rural or inner-city students in control group that are not sampled in experimental sample, sample $n_{\text{obs}}/4$ rural or inner-city students in treatment group whose year 1 average grade is in the lower half among treated rural or inner-city students, sample $n_{\text{obs}}/4$ urban or suburban students in control group and finally sample $n_{\text{obs}}/4$ urban or suburban students in treatment group whose year 1 average grade is in lower half among treated urban or suburban students. This is our observational sample (which is confounded because we remove students with higher scores selectively from the population) $\mathcal{D}_O$.
    \item Use different methods to estimate $\tau_{gt}$ based on $\mathcal{D}_E$ and $\mathcal{D}_O$.
    \item Compare based on mean squared error (MSE).
\end{enumerate}

\begin{table} 
\centering
\scalebox{1.0} {
\begin{tabular}{c l c c c c} 
\toprule 
$n_{\text{exp}}$ & $n_{\text{obs}}$ & GRF & Imputation & AIPW\\
\midrule
    \multirow{1}{*}{300} 
             & 1000 & 7.08 & 13.19 & 167.52\\
            \midrule
    \multirow{1}{*}{200} 
             & 1500 & 9.36 & 12.76 & 167.43\\
            \midrule
    \multirow{1}{*}{500} 
             & 2000 & 4.54 & 7.43 & 166.08\\
\bottomrule 
\end{tabular}}
\caption{STAR study simulation} 
\label{table:star1} 
\end{table}

\begin{table} 
\centering
\scalebox{1.0} {
\begin{tabular}{c l c c c c} 
\toprule 
$n_{\text{exp}}$ & $n_{\text{obs}}$ & GRF & Imputation & AIPW & $\tau$\\
\midrule
    \multirow{1}{*}{300} 
             & 1000 & 6.64 & 7.90 & -5.21 & 7.62\\
            \midrule
    \multirow{1}{*}{200} 
             & 1500 & 6.89 & 8.21 & -5.24 & 7.62\\
            \midrule
    \multirow{1}{*}{500} 
             & 2000 & 6.70 & 8.06 & -5.21 & 7.62\\
\bottomrule 
\end{tabular}}
\caption{STAR study simulation, empirical mean and true treatment effect} 
\label{table:star_mean} 
\end{table}

We will only compare GRF and imputation estimator as the Kallus method involves estimating the coefficient of a linear function of the covariates but we only have categorical variables. We also include the mean squared error of the AIPW estimator (notice that AIPW estimator requires the sample to be unconfounded) on observational sample. Table~\ref{table:star1} gives the results. We see that in general the GRF estimator outperforms the imputation estimator and these two estimators all outperform the AIPW estimator significantly. In particular, as Table~\ref{table:star_mean} shows, the empirical mean of AIPW estimates is actually a negative number (and the true treatment effect is a positive number) and is far from the true treatment effect.

\section{Conclusion}
In this paper, we proposed a simple procedure to estimate the average treatment effect of the primary outcome in observational study by utilizing an experimental study for the surrogate outcome. We showed that our procedure can be applied in many settings so long as we can estimate the conditional average treatment effect of the surrogate outcome. We compared several methods through simulations and showed that our procedure gives better estimate in terms of mean square error than the canonical imputation estimator in \cite{athey2020}.

\section{Acknowledgement}
The author thanks Kevin Guo and Guido Imbens for valuable discussions.

\bibliographystyle{abbrvnat}
\bibliography{main}

\end{document}